\documentclass[a4paper,aps,pre,11pt]{revtex4}

\usepackage{epsfig,amssymb,graphicx}

\begin{document}

\title{Linear Quantum Entropy and Non-Hermitian Hamiltonians}

\author{Alessandro Sergi}
\email{asergi@unime.it}

\affiliation{
Dipartimento di Scienze Matematiche e Informatiche, \\
Scienze Fisiche e Scienze della Terra, \\
Universit\`a degli Studi di Messina, \\
Contrada Papardo, 98166 Messina, Italy}

\affiliation{Institute of Systems Science,
Durban University of Technology,
P. O. Box 1334, Durban 4000, South Africa}

\author{Paolo V. Giaquinta}
\email{Paolo.Giaquinta@unime.it}
\affiliation{School of Physics and National Institute for
Theoretical Physics, University of KwaZulu-Natal, Westville Campus,
Private Bag X54001, Durban 4000, South Africa}

\begin{abstract}
We consider the description of open quantum systems with probability sinks (or sources) in terms of general non-Hermitian Hamiltonians.~Within such a framework, we study {novel}
 possible definitions of the quantum linear entropy as an indicator of the flow of information
during the dynamics. {Such linear entropy functionals are necessary
in the case of a partially Wigner-transformed non-Hermitian Hamiltonian (which is
typically useful within a mixed quantum-classical representation)}.
Both the case of a system represented by a pure non-Hermitian Hamiltonian as well as that of
the case of non-Hermitian dynamics in a classical bath are explicitly considered.
{\bf Published in \emph{Entropy} 18, 451 (2016); doi:10.3390/e18120451}.
\end{abstract}

\maketitle

\section{Introduction}

The study of open quantum systems is one of the fundamental problems
of modern physics~\cite{rhbook,bpbook}.
An open quantum system consists of a region of space where
quantum processes take place (and which can be studied by the
experimenter) in contact with a decohering and dissipative
environment that is typically beyond the control of the experimenter.
Various instances of concrete open quantum systems can be found
in different areas of physics such as, for example, quantum optics,
atomic and mesoscopic physics, biophysics or, at even shorter
distances, nuclear physics.
The interdisciplinary character of the theory of open quantum systems
calls for a variety of different approaches.
Here,~we are concerned in particular with a formalism that adopts
non-Hermitian Hamiltonian operators, a~theoretical approach that
is routinely called non-Hermitian quantum mechanics~\cite{moyeseyev}.
The description of open quantum systems in terms of non-Hermitian
Hamiltonians~\cite{rotter-bird} can be rigorously derived, in~the case
of a localised quantum subsystem coupled to a continuum of scattering
states, by means of the
Feshbach projection formalism~\cite{fesh,fesh2,rotter}.
{Such an approach has been successfully employed to illustrate
the complexities of exceptional points, which do occur
when resonances coalesce in a non-avoided
crossing~\cite{rotter-bird}.
When one uses the full non-Hermitian Hamiltonian,
left and right
eigenvectors~\cite{eleuch-rotter,eleuch-rotter2,eleuch-rotter3}
must be distinguished.
From this perspective, the occurrence of exceptional points
may create problems for defining the density matrix. 
On the other side, one can always use a Hermitian basis
(which, for~example, but not necessarily,
arises from the Hermitian part of the full
non-Hermitian operators)
to represent non-Hermitian operators and the density matrix.
From such a vantage point, the coalescence of the eigenvalues
of the non-Hermitian Hamiltonian appears to be a foregoer of
such major problems.}
It is worth mentioning that non-Hermitian Hamiltonians also appear
in parity-time (PT) symmetric generalisations of
quantum mechanics~\cite{bender,mostafazadeh}.
Such {theories} have recently found concrete applications
in lossy optical waveguides~\cite{optics,optics2}
and photonic lattices~\cite{lattice,lattice2}.

However, we are interested here in open systems that can be
effectively described by non-Hermitian Hamiltonians that are
not necessarily PT-symmetric (and which, for such a reason, will
be called general in the rest of this paper).
For such Hamiltonians, it has been shown how to define
a proper statistical mechanics~\cite{ks} in order to study the
behaviour of non equilibrium averages (e.g., the
purity of quantum states~\cite{kostya-purity}) and to provide the
definition of correlation functions~\cite{sk-corr}.

In order to try to build 
possible measures of quantum 
information~{\cite{mahler,mahler2,qinfo}}
for systems with
general non-Hermitian Hamiltonians, one can start
by defining an entropy functional~\cite{vonneumann,ohya}.
To~this end, a~non-Hermitian generalisation
of the von Neumann entropy has been introduced in \cite{sk-S}.
Nevertheless,~entropies of the von Neumann form cannot be used
when quantum theory is formulated by means of the Wigner function~\cite{manfredi}.
Since the (partial) Wigner representation is particularly useful
in order to derive a mixed quantum-classical description of
non-Hermitian systems~\cite{nhps}, it becomes interesting to
study the properties of the
so-called linear entropy~\cite{manfredi,zurek,pattanayak}
and its generalisation to the case of open quantum systems
described by general non-Hermitian Hamiltonians.
{To this end, we~present in this paper, 
for the first time to our
knowledge, a generalisation of the entropy for
systems with non-Hermitian Hamiltonians that must be adopted
when there is an embedding of the quantum subsystem in phase space.
We associate the term ``linear'' to such an 
entropy as it arises from its first appearance 
in the literature~\cite{manfredi,zurek,pattanayak}.}

This paper is organised as follows. In Section~\ref{sec:s-nh},
we summarise the results of the
density-matrix approach~\cite{ks,kostya-purity,sk-corr,sk-S}
to non-Hermitian dynamics that are useful for the study and
generalisation of the linear entropy~\cite{manfredi,zurek,pattanayak}.
In particular, we introduce the equations of motion
for the density matrices~\cite{ks} and the von Neumann-like entropies
studied in \cite{sk-S}.
In Section~\ref{sec:qlinS}, we study the linear entropy
and its non-Hermitian generalisation, along the lines
followed in \cite{sk-S}.
Analytical solutions are given in the
{basic} case of
a constant decay operator.
{It is worth noting that even basic models with 
constant decay operators become interesting when one adds
the additional level of complexity provided by the classical-like
environment represented by means of the partially Wigner-transformed
Hermitian part of the Hamiltonian. In order to fix the ideas,
one can think of a light-emitting quantum dot coupled to
an energy-absorbing optical guide in a classical environment,
which introduces thermal fluctuations or some other type of noise.
It is not even difficult to imagine how models like these one can be
made more and more complex within our approach.}
In Section~\ref{sec:qcnh}, we briefly recall how to formulate
the dynamics of a non-Hermitian system that is embedded in
a classical bath of degrees of freedom.
In Section~\ref{sec:qcentropy}, we study the behaviour of
the linear entropy and its non-Hermitian generalisation
in a quantum-classical set-up. Once again,
analytical solutions are provided for the case
of a constant decay operator.
Finally, our conclusions are presented in Section~\ref{sec:concl}.

\section{Quantum Dynamics with Non-Hermitian Hamiltonians}
\label{sec:s-nh}

Let us consider a non-Hermitian Hamiltonian composed of two terms:
\begin{equation}
\hat{\cal H}=
\hat H -i \hat\Gamma \;.
\label{eq:tot-NH-H}
\end{equation}

Both operators on the right-hand side,
$\hat H$ and $\hat\Gamma$, are Hermitian; 
$\hat\Gamma$ is often called the decay rate~operator.
The quantum states $|\Psi\rangle$ and $\langle\Psi|$
evolve according to the Schr\"odinger equations
\begin{eqnarray}
\partial_t|{\Psi}\rangle&=&-\frac{i}{\hbar}\hat{\cal H}|\Psi\rangle=
-\frac{i}{\hbar}\hat H|\Psi\rangle-\frac{1}{\hbar}\hat\Gamma|\Psi\rangle
\;, \\
\partial_t \langle {\Psi}|&=&\frac{i}{\hbar}\langle\Psi |\hat{\cal H}^\dagger
=\frac{i}{\hbar}\langle\Psi|\hat H-\frac{1}{\hbar}\langle\Psi|\hat\Gamma
.
\end{eqnarray}

On conceptual grounds, we can expect that the open quantum system
dynamics produces statistical mixtures. Indeed, 
we have shown that the purity is not conserved \cite{ks,kostya-purity}.
Defining the non-normalised density matrix as
\begin{equation}
\hat{\Omega} =\sum_k{\cal P}_k|\Psi^k \rangle\langle\Psi^k |
\;,
\end{equation}
{where 
$(|\Psi^k \rangle,\langle\Psi^k |)$  are the eigenstates of any good
\emph{Hermitian} operator that can cover the Hilbert space
of the system and
${\cal P}_k$ is their} probability of occurrence,
the equation of motion can be written as
\begin{equation}
\partial_t {\hat{\Omega}} =-\frac{i}{\hbar}\left[\hat H, \hat{\Omega}\right]_- -
\frac{1}{\hbar}
\left[\hat{\Gamma},\hat{\Omega} \right]_+
\;,
\label{eq:dotOmega}
\end{equation}
with $[~,~]_-$ and $[~,~]_+$ denoting the commutator
and anticommutator, respectively.~Equation (\ref{eq:dotOmega}) effectively describes the subsystem
(with Hamiltonian $\hat{H}$) coupled to the environment (represented
by $\hat{\Gamma}$). \linebreak
{It~is worth remarking again and explicitly that, in our
approach~\cite{ks,sk-corr,sk-S,nhps}, we use 
Hermitian basis sets to represent the equations of motion.
This situation is commonly found when, for example,
the~non-Hermitian creation and destruction operators, $\hat{a}$ 
and $\hat{a^\dag}$, respectively, are represented in the
basis of the Hermitian number operator.
It should be evident that, because of this, we do not need to worry about
the left/right eigenvectors of the full 
non-Hermitian Hamiltonian~\cite{leftright1,leftright2}.
} 

Non-Hermitian dynamics do not conserve the probability.
This can be easily seen by
taking the trace of both sides of Equation~(\ref{eq:dotOmega}):
\begin{equation}
\partial_t{\rm Tr}\,\hat{\Omega}
=-\frac{2}{\hbar}{\rm Tr}\left(\hat{\Gamma} \,\hat{\Omega} \right)
\;.
\label{eq:dotTrOmega}
\end{equation}

However, we can define a normalised density matrix~\cite{ks} as
\begin{equation}
\hat{\rho} =\frac{\hat{\Omega} }{{\rm Tr}\,\hat{\Omega} } \;.
\label{eq:rho}
\end{equation}

The density matrix in Equation (\ref{eq:rho})
can be used in the calculation of statistical averages: \linebreak
$\langle\chi\rangle_t={\rm Tr} \left( \hat\chi\hat{\rho}(t) \right),$
where $\hat\chi$ is an arbitrary operator. 
The normalised density matrix $\hat\rho$
obeys the equation~\cite{ks}:
\begin{equation}
\partial_t {\hat{\rho}} =
-\frac{i}{\hbar}\left[\hat H, \hat{\rho}\right]_-
-\frac{1}{\hbar}\left[\hat{\Gamma},\hat{\rho}\right]_+
+\frac{2}{\hbar}\hat{\rho} \, {\rm Tr} (\hat{\Gamma}\hat{\rho})\;.
\label{eq:dotrho}
\end{equation}

Similarly to Equation (\ref{eq:dotOmega}),
Equation (\ref{eq:dotrho}) effectively describes the evolution of 
the subsystem coupled to the environment;
the role of the third term on the right-hand side is to
conserve the probability during the dynamics.
Equation~(\ref{eq:dotrho}) is nonlinear.
This property was also noted when considering
operator averages in \cite{emg-korsch}).
Within the Feshbach--Fano projection formalism,
the nonlinearity of the non-Hermitian approach
has been suggested in~\cite{zno02} as well.
While the density operator $\hat{\rho}$
is bounded and useful in the calculation of
of the statistical averages,
the gain or loss of probability of open systems
are properly described
by the non-normalised density operator $\hat{\Omega}$.
Hence, it turns out that both $\hat{\Omega}$ and $\hat{\rho}$
are useful in the non-Hermitian formalism~\cite{sk-corr,sk-S}.

The normalised density matrix $\hat\rho$ allows us to
define~\cite{sk-S}
the von Neumann entropy of a non-Hermitian system as
\begin{equation}
S_{\rm vN} \equiv -k_B \left\langle   \ln\hat{\rho} \right\rangle
= -k_B{\rm Tr}\left( \hat{\rho} \ln\hat{\rho} \right)
\; .
\label{eq:SVNnH}
\end{equation}

The rate of the von Neumann entropy production is~\cite{sk-S}:
\begin{eqnarray}
\partial_t S_{\rm vN}
= 
\frac{2k_B}{\hbar}{\rm Tr}\left(\hat{\Gamma}\hat{\rho} \ln\hat{\rho} \right)
+
\frac{2}{\hbar}{\rm Tr}\left( \hat{\Gamma}\hat{\rho} \right)
S_{\rm vN}
. \label{eq:dotSVNnH}
\end{eqnarray}

However, the gain or loss of information in a non-Hermitian system
are more properly represented by introducing
another entropy, given by
the statistical average of the logarithm
of the \emph{non-normalised} density operator~\cite{sk-S}:
\begin{equation}
S_{\rm NH} \equiv -k_B \langle   \ln\hat{\Omega} \rangle
= -k_B{\rm Tr} ( \hat{\rho} \ln\hat{\Omega} ) 
= - k_B \frac{ {\rm Tr}( \hat{\Omega} \ln\hat{\Omega} ) }{
{\rm Tr}\, \hat{\Omega} } \; .
\label{eq:SVNnH2}
\end{equation}

The rate of change of $S_{\rm NH}$ is~\cite{sk-S}
\begin{equation}
\partial_t S_{\rm NH}
=\frac{2k_{\rm B}}{\hbar}{\rm Tr}\left(\hat{\Gamma}\hat{\rho}\ln\hat{\Omega}\right)
+\frac{2}{\hbar} {\rm Tr}\left(\hat\Gamma \rho\right) S_{\rm NH}
+2\frac{k_B}{\hbar}{\rm Tr}\left(\hat{\Gamma}\hat{\rho}\right)\;,
\end{equation}
while the difference between the two entropies reads
\begin{equation}
S_{\rm vN} - S_{\rm NH} = k_{\rm B}
\ln\left( {\rm Tr}\, \hat\Omega\right)\; .
\label{e:diffentr}
\end{equation}

The fact that the $S_{\rm NH}$ entropy captures the
expected physical behaviour of the flow of information
out of an open system can be seen by considering
the models where $\hat H$ is an arbitrary self-adjoint
operator while $\hat \Gamma$ is proportional to the 
identity operator:
\begin{equation}
\hat\Gamma = \frac{1}{2} \hbar \gamma_0 \hat I \;,
\label{eq:Gammamodgau}
\end{equation}
where the parameter $\gamma_0$ is assumed to be real-valued.
For such models, after imposing the initial conditions 
${\rm Tr}\,\hat{\Omega} (0) = 1$),
we obtain~\cite{sk-S}:
\begin{eqnarray}
{\rm Tr}\,\hat{\Omega} (t) &=& \exp{(-\gamma_0 t)}\;,\label{eq:trOmega}\\
S_{\rm vN} (t) &=& S_{\rm vN}^{(0)} = \text{const}\;,\\
S_{\rm NH} (t) &=& S_{\rm vN}^{(0)} + k_{\rm B} \gamma_0 t\;.
\end{eqnarray}

One can then see that, for positive values of $\gamma_0$,
the $S_{\rm NH}$ entropy diverges 
at large times, as 
{a good entropy functional}
of an open system is expected to do.
On the contrary, the von Neumann entropy $S_{\rm vN}$ 
is always constant.

\section{Non-Hermitian Dynamics and Quantum Linear Entropy}
\label{sec:qlinS}

The quantum linear entropy is
\begin{equation}
S_{\rm lin}=1-{\rm Tr}\left[\hat{\rho}^2(t)\right].
\label{eq:Slin}
\end{equation}

The entropy production is
\begin{equation}
\dot{S}_{\rm lin}
=-2{\rm Tr}\left[\hat{\rho}(t)\frac{\partial\hat{\rho}(t)}{\partial t}
\right]\;.
\label{eq:dotSlin-ini}
\end{equation}

Substituting Equation~(\ref{eq:dotrho}) in Equation~(\ref{eq:dotSlin-ini})
and using the following identities
\begin{eqnarray}
{\rm Tr}\left[\hat{\rho}\hat{H}\hat{\rho}-\hat{\rho}\hat{\rho}\hat{H}\right]
&=& 
{\rm Tr}\left[\hat{\rho}^2\hat{H}-\hat{\rho}^2\hat{H}\right]=0\;,
\\
{\rm Tr}\left[\hat{\rho}\hat{\Gamma}\hat{\rho}+\hat{\rho}\hat{\rho}\hat{\Gamma}
\right]
&=&
2{\rm Tr}\left[\hat{\Gamma}\hat{\rho}^2\right] \;, 
\end{eqnarray}
we obtain
\begin{equation}
\dot{S}_{\rm lin}
=\frac{4}{\hbar}{\rm Tr}\left[\hat{\Gamma}\hat{\rho}^2(t)\right]
-\frac{4}{\hbar}{\rm Tr}\left[\hat{\Gamma}\hat{\rho}(t)\right]
{\rm Tr}\left[\hat{\rho}^2(t)\right]\;.
\label{eq:dotSlin}
\end{equation}

Analogously with the entropy of Equation~(\ref{eq:SVNnH2}),
we can also introduce a linear entropy involving the
non-normalised density matrix as
\begin{equation}
S_{\rm lin}^{\rm NH}=1-{\rm Tr}\left[\hat{\rho}(t)\hat{\Omega}(t)\right]
\;.
\label{eq:SlinNH}
\end{equation}

The rate of production of $S_{\rm lin}^{\rm NH}$ is
\begin{eqnarray}
\dot{S}_{\rm lin}^{\rm NH}
&=&
-\frac{2}{{\rm Tr}\left[\hat{\Omega}(t)\right]}
{\rm Tr}\left[\hat{\Omega}(t)\partial_t\hat{\Omega}(t)\right]
-
\frac{2{\rm Tr}\left(\hat{\Omega}^2(t)\right)}
{\hbar\left[{\rm Tr}\left(\hat{\Omega}(t)\right)\right]^2}
{\rm Tr}\left(\hat{\Gamma}\hat{\Omega}(t)\right)
\;.\label{eq:dotSNHlin-ini}
\end{eqnarray}

Using Equation~(\ref{eq:dotOmega}), together with the identity
\begin{eqnarray}
{\rm Tr}\left[\hat{\Omega}[\hat{\Gamma},\hat{\Omega}]_+\right]
&=&
2{\rm Tr}\left[\hat{\Gamma}\hat{\Omega}^2\right]\;,
\end{eqnarray}
in Equation (\ref{eq:dotSNHlin-ini}), we obtain
\begin{eqnarray}
\dot{S}_{\rm lin}^{\rm NH}
&=&
\frac{4{\rm Tr}\left[\hat{\Gamma}\hat{\Omega}^2(t)\right]}
{\hbar{\rm Tr}\left[\hat{\Omega}(t)\right]}
-
\frac{2{\rm Tr}\left(\hat{\Omega}^2(t)\right)
{\rm Tr}\left(\hat{\Gamma}\hat{\Omega}(t)\right)
}
{\hbar\left[{\rm Tr}\left(\hat{\Omega}(t)\right)\right]^2}
\;.
\label{eq:dotSlinNH}
\end{eqnarray}

\subsection*{Linear Entropy Production and Constant Decay Operator}

Let us consider Equations (\ref{eq:dotSlin}) and~(\ref{eq:dotSlinNH})
in the case of a decay operator defined by Equation~(\ref{eq:Gammamodgau}).
In~such a case, the temporal dependence of ${\rm Tr}(\hat\Omega(t))$
is given, when choosing ${\rm Tr}\hat\Omega(0)=1$, 
by Equation~(\ref{eq:trOmega}).
Using Equation~(\ref{eq:dotOmega}), we easily obtain
\begin{eqnarray}
\partial_t {\rm Tr} \hat \Omega^2(t)
&=& -2{\gamma_0}{\rm Tr}\hat\Omega^2(t),
\\
{\rm Tr} \hat \Omega^2(t)
&=&
{\rm Tr} \hat \Omega^2(0)\exp[-2{\gamma_0} t]
\;.
\end{eqnarray}

Hence, the calculation of
\begin{equation}
\partial_t {\rm Tr} \hat \rho^2(t) = 2 {\rm Tr}\left[\hat\rho(t)
\partial_t\hat\rho(t)\right] \;
\label{eq:dotTrrho2-ini}
\end{equation}
can proceed upon considering the identities
\begin{eqnarray}
-\frac{2}{\hbar}{\rm Tr}
\left\{\hat\rho(t)\left[\hat\Gamma,\hat\rho(t)\right]_+\right\}
&=&-2{\gamma_0}{\rm Tr}\left[\hat\rho^2(t)\right]
\;,\\
\frac{4}{\hbar}{\rm Tr}\left\{\hat\rho^2(t)
{\rm Tr}\left[\hat\Gamma\hat\rho(t)\right]\right\}
&=&
2{\gamma_0} {\rm Tr}\left[\hat\rho^2(t)\right] \;.
\end{eqnarray}

Therefore, Equation~(\ref{eq:dotTrrho2-ini}) is found to give

\begin{equation}
\partial_t{\rm Tr}\hat\rho^2(t)
=-2{\gamma_0} {\rm Tr}\left[\hat\rho^2(t)\right]
+2{\gamma_0} {\rm Tr}\left[\hat\rho^2(t)\right]=0\;.
\end{equation}

Given the above result, we can choose

\begin{equation}
{\rm Tr}\hat\rho^2(t)={\rm const.}={\rm Tr}\hat\rho^2(0) \;.
\end{equation}

Finally, Equation~(\ref{eq:dotSlin}) becomes
\begin{eqnarray}
\dot{S}_{\rm lin}
&=&
2{\gamma_0}{\rm Tr}[\hat\rho^2(0)]
-2{\gamma_0}{\rm Tr}[\hat\rho^2(t)]=0\;.
\label{eq:dotSlingamma-const}
\end{eqnarray}

Equation~(\ref{eq:dotSlingamma-const}) shows that $S_{\rm lin}$
is identically constant and is thus not suitable
to describe the information flow or the evolution
of the entanglement in systems with non-Hermitian Hamiltonians.

Let us now consider Equation~(\ref{eq:dotSlinNH}): it becomes
\begin{eqnarray}
\dot{S}_{\rm lin}^{\rm NH}
&=&
2{\gamma_0}\frac{{\rm Tr}\hat\Omega^2(t)}{{\rm Tr}\hat\Omega(t)}
-{\gamma_0}\frac{{\rm Tr}\hat\Omega^2(t)}{{\rm Tr}\hat\Omega(t)}
\nonumber\\
&=&
{\gamma_0}\left[{\rm Tr}\hat\Omega^2(0)\right]e^{-\gamma t}\;.
\end{eqnarray}

Integrating between $0$ and $t$, we obtain
\begin{equation}
S_{\rm lin}^{\rm NH}=\left[1-e^{-{\gamma_0}t}\right]
{\rm Tr} \hat\Omega^2(0)
\;.\label{eq:SlinNH-t}
\end{equation}

Equation~(\ref{eq:SlinNH-t}) describes the increase of the linear entropy
$S_{\rm lin}^{\rm NH}$ from the value of $0$ at $t=0$
to the plateau value of ${\rm Tr} \hat\Omega^2(0)$ at $t=\infty$.
Because of the choice of the initial condition ${\rm Tr} \hat\Omega(0)=1$,
the~quantity ${\rm Tr} \hat\Omega^2(0)$ is the purity of the
non-Hermitian system. Hence, Equation~(\ref{eq:SlinNH-t}) monitors the loss
of the initial purity of the system.

\section{Non-Hermitian Dynamics in a Classical Environment}
\label{sec:qcnh}

One particular class of open quantum systems is obtained when
a quantum subsystem is embedded in a classical environment.~In \cite{nhps}, an equation of motion for a quantum subsystem
embedded in~a~classical bath, described in terms of its phase space
coordinates, has been derived.
To~this end, we consider a~total~Hamiltonian

\begin{equation}
\hat{\cal H}(\hat{r},\hat{p},\hat{R},\hat{P})
=\hat{H}(\hat{r},\hat{p},\hat{R},\hat{P})
-i\hat{\Gamma}(\hat{r},\hat{p}) \;,
\label{eq:calH}
\end{equation}
where $(\hat{r},\hat{p})$ are $n$ light degrees of freedom
with mass $m$, and $(\hat{R},\hat{P})$ are $N$ heavy degrees of
freedom of mass $M$. The small expansion parameter $\mu=\sqrt{m/M}
<<1$ can be used to obtain the classical limit for the 
$(\hat{R},\hat{P})$ degrees of freedom, after taking
a partial Wigner transform over the $2N$ heavy coordinates.
Using a multidimensional notation and denoting the phase space point
$(R,P)$ with $X$, the~partial Wigner transform
of the density matrix is defined as

\begin{equation}
\hat{\Omega}_{\rm W}(X,t)
=\frac{1}{(2\pi\hbar)^N}\int dZe^{P\cdot Z/\hbar}
\langle R-Z/2|\hat{\Omega}(t)|R+Z/2\rangle
\;,
\end{equation}
while the partial Wigner transform of an arbitrary
operator $\hat{\chi}$ is defined as

\begin{equation}
\hat{\chi}_{\rm W}(X)
=\int dZe^{P\cdot Z/\hbar}
\langle R-Z/2|\hat{\chi}|R+Z/2\rangle
\;.
\end{equation}

In \cite{nhps}, it was shown that, upon taking the
partial Wigner transform of Equation~(\ref{eq:dotOmega}),
with the $\hat{H}$ and $\hat{\Gamma}$ of Equation~(\ref{eq:calH}),
and performing a linear expansion in $\mu$, one obtains
the equation of motion
\begin{eqnarray}
\frac{\partial}{\partial t}\hat{\Omega}_{\rm W}(X,t)
&=&-\frac{i}{\hbar}
\left[\hat{H}_{\rm W},\hat{\Omega}_{\rm W}(X,t)\right]_-
+
\frac{1}{2} 
{\cal B}_{ab} \left(\partial_a\hat{H}_{\rm W}\right)
\left(\partial_b\hat{\Omega}_{\rm W}(X,t)\right)
\nonumber\\
&-&
\frac{1}{2}
{\cal B}_{ab}\left(\partial_a\hat{\Omega}_{\rm W}(X,t)\right)
\left(\partial_b\hat{H}_{\rm W}\right)
-
\frac{1}{\hbar}\left[\hat{\Gamma},\hat{\Omega}_{\rm W}(X,t)\right]_+
\;, \label{eq:qcdotOm}
\end{eqnarray}
where ${\cal B}_{ab}=-{\cal B}_{ba}^{\rm T}$
is the symplectic matrix~\cite{mccauley}
and $\partial_a=(\partial/\partial X_a)$ is
the gradient operator in phase space.
The Einstein convention of summing over repeated indices
is used throughout this paper.
One~can note that ${\cal B}_{ab} (\partial_a\hat{H}_{\rm W})
(\partial_b\hat{\Omega}_{\rm W})$
is the Poisson bracket between $ \hat{H}_{\rm W}$ and 
$\hat{\Omega}_{\rm W}$.

Equation~(\ref{eq:qcdotOm}) describes the evolution of the
non-normalised density matrix, $\hat{\Omega}_{\rm W}(X,t)$,
when a quantum subsystem with probability sinks or sources
(represented by the decay operator $\hat{\Gamma}$)
is embedded in a classical environment (with phase space coordinates
$X$). The classical bath produces both statistical noise and 
 decoherence in addition to those eventually represented by the
decay operator.
As a consequence of Equation~(\ref{eq:qcdotOm}), the trace of
$\hat{\Omega}_{\rm W}(X,t)$ is not a conserved quantity:
\begin{eqnarray} 
\frac{d}{dt}{\rm Tr}^\prime\int dX \hat{\Omega}_{\rm W}(X,t)
&=&
\frac{d}{dt}\tilde{\rm T}{\rm r}\left[ \hat{\Omega}_{\rm W}(X,t)\right]
\nonumber\\
&=&
\tilde{\rm T}{\rm r}\left[ \frac{\partial }{\partial t}\hat{\Omega}_{\rm W}(X,t)\right]
\neq 0\;,
\end{eqnarray} 
where we have denoted with the symbol ${\rm Tr}^\prime$ a partial trace
over the quantal degrees of freedom, with the symbol $\int dX$ the phase
space integral, and with the symbol $\tilde{\rm T}{\rm r}$ both the
partial trace and the phase space integral.

Using the cyclic invariance of the trace, we can easily see that
\begin{eqnarray}
\tilde{\rm T}{\rm r}\left\{
\left[\hat{H}_{\rm W},\hat{\Omega}_{\rm W}\right]_-
\right\} 
&=&
\tilde{\rm T}{\rm r}\left\{
\hat{H}_{\rm W}\hat{\Omega}_{\rm W}-\hat{H}_{\rm W}\hat{\Omega}_{\rm W}
\right\} =0\;,
\\
\tilde{\rm T}{\rm r}\left\{
\hat{H}_{\rm W}\overleftarrow{\nabla}_a 
{\cal B}_{ab}
\overrightarrow{\nabla}_b \hat{\Omega}_{\rm W}
-
\hat{\Omega}_{\rm W} \overleftarrow{\nabla}_a
{\cal B}_{ab}
\overrightarrow{\nabla}_b \hat{H}_{\rm W}
\right\}
&=& 0 \;,
\end{eqnarray}
where, in the last identity, we have also performed an
integration by parts and exploited
the fact that ${\cal B}_{ab}$
are constants.
If we also use the identity
\begin{eqnarray}
\tilde{\rm T}{\rm r}\left\{
\left[\hat{\Gamma},\hat{\Omega}_{\rm W}\right]_+
\right\}
&=&
2{\rm Tr}^\prime\left[\hat{\Gamma}\hat{\Omega}_{\rm S}\right]
\;,
\end{eqnarray}
where $\hat{\Omega}_{\rm S}=\int dX \hat{\Omega}_{\rm W}(X)$,
we can then find
\begin{equation}
\frac{d}{dt}\tilde{\rm T}{\rm r}\left[\hat{\Omega}_{\rm W}(X,t)\right]
=
-\frac{2}{\hbar}
{\rm Tr}^\prime\left[\hat{\Gamma}\hat{\Omega}_{\rm S}(t)\right]
\;.
\label{eq:dotqcTrace}
\end{equation}

Equation~(\ref{eq:dotqcTrace}) is analogous to Equation~(\ref{eq:dotTrOmega})
and shows that the probability is not conserved
for the quantum-classical system because of the action of
the decay operator.
We can introduce a normalised density matrix as

\begin{equation}
\hat{\rho}_{\rm W}(X,t)=\frac{\hat{\Omega}_{\rm W}(X,t)}
{\tilde{\rm T}{\rm r}\left[ \hat{\Omega}_{\rm W}(X,t)\right]} \;,
\label{eq:rhoW}
\end{equation}
and, using Equations~(\ref{eq:qcdotOm}) and~(\ref{eq:dotqcTrace}),
find its equation of motion:
\begin{equation}
\begin{array}{ll}
\frac{\partial}{\partial t}\hat{\rho}_{\rm W}(X,t)
=-\frac{i}{\hbar}
\left[\hat{H}_{\rm W},\hat{\rho}_{\rm W}(X,t)\right]_-
+
\frac{1}{2} \hat{H}_{\rm W}\overleftarrow{\nabla} \cdot
\mbox{\boldmath${\cal B}$} \cdot
\overrightarrow{\nabla} \hat{\rho}_{\rm W}(X,t)\vspace{5pt} \\
~~~~~~~~~~~~~~~~~~~~~~~-\frac{1}{2} \hat{\rho}_{\rm W}(X,t) \overleftarrow{\nabla} \cdot
\mbox{\boldmath${\cal B}$} \cdot
\overrightarrow{\nabla} \hat{H}_{\rm W} \vspace{5pt} \\
~~~~~~~~~~~~~~~~~~~~~~~-\frac{1}{\hbar}\left[\hat{\Gamma},\hat{\rho}_{\rm W}(X,t)\right]_+
+\frac{2}{\hbar}\hat{\rho}_{\rm W}(X,t)
\tilde{\rm T}{\rm r}\left[\hat{\Gamma}\hat{\rho}_{\rm W}(X,t)\right]
\;. 
\label{eq:qcdotrhoW}
\end{array}
\end{equation}

At variance with Equation~(\ref{eq:qcdotOm}), Equation~(\ref{eq:qcdotrhoW})
is nonlinear and allows one to define averages of the dynamical
variables of the quantum-classical system with a non-Hermitian
Hamiltonian that has a probabilistic meaning.

\section{Entropy Production and Quantum-Classical Non-Hermitian
Hamiltonians}
\label{sec:qcentropy}

As noted in~\cite{manfredi}, when considering the definition of
the entropy for a quantum system in terms of the Wigner function,
the typical choice in terms of the von Neumann definition,
found in Equation~(\ref{eq:SVNnH}) when the Wigner function
$f_{\rm W}(x,X,t)$ replaces the density matrix $\hat\rho$,
cannot work: $f_{\rm W}(x,X,t)$ can be negative in general.
What one can do~\cite{manfredi} is start from the linear
entropy~\cite{zurek,pattanayak}, $S_{\rm lin}=1-{\rm Tr}(\hat{\rho}^2)$,
and perform the Wigner transform in order to obtain:
\begin{equation}
S_{\rm lin}=1-(2\pi\hbar)^{n+N}\int dxdX f_{\rm W}^2(x,X,t)\;,
\label{eq:s2}
\end{equation}
where $f_{\rm W}(x,X,t)$ is the Wigner function, obtained
by transforming $\hat{\rho}$ over all the coordinates.

In a mixed quantum-classical framework, the natural extension
of Equation~(\ref{eq:s2}) is given by
\begin{equation}
S_{\rm lin,W}=1-(2\pi\hbar)^{N}{\rm Tr}^\prime\int dX\hat{\rho}_{\rm W}^2(X,t)
=1-(2\pi\hbar)^{N}\tilde{\rm T}{\rm r}\left[\hat{\rho}_{\rm W}^2(X,t)\right].
\label{eq:s2W}
\end{equation}

When considering the non-Hermitian dynamics of the quantum subsystem 
embedded in the classical environment, given by Equation~(\ref{eq:qcdotrhoW}),
we obtain the linear entropy production
\begin{equation}
\dot{S}_{\rm lin,W}=
-2(2\pi\hbar)^{N}\tilde{\rm T}{\rm r}\left[\hat{\rho}_{\rm W}
\frac{\partial\hat{\rho}_{\rm W}}{\partial t}\right].
\label{eq:s2pW}
\end{equation}

{We have obtained 
Equation~(\ref{eq:s2pW}) by using the identities
\begin{eqnarray}
\tilde{\rm T}{\rm r}\left\{\hat{\rho}_{\rm W}
\left[\hat{H}_{\rm W},\hat{\rho}_{\rm W}\right]_-
\right\}&=&0\;,
\\
\tilde{\rm T}{\rm r}\left\{\hat{\rho}_{\rm W}
\left[{\cal B}_{ab}(\nabla_a\hat{H}_{\rm W})(\nabla_b\hat{\rho}_{\rm W})
-
{\cal B}_{ab}(\nabla_a\hat{\rho}_{\rm W})(\nabla_b\hat{H}_{\rm W})
\right]
\right\}
&=&0\;,
\end{eqnarray}
 together with}
${\cal B}_{ab} (\nabla_{ab}^2\hat{\rho}_{\rm W}) \hat{\rho}_{\rm W}
\hat{H}_{\rm W}=0$ and ${\cal B}_{ab} \hat{\rho}_{\rm W}
(\nabla_{ab}^2\hat{\rho}_{\rm W}) \hat{H}_{\rm W}=0$,
which follow from taking the trace of an antisymmetric matrix, ${\cal B}_{ab}$,
and a symmetric one, $\nabla_{ab}^2\hat{\rho}_{\rm W}$.
Noting that we also have
\begin{eqnarray}
\tilde{\rm T}{\rm r}\left\{\hat{\rho}_{\rm W}
\left[\hat{\Gamma},\hat{\rho}_{\rm W}\right]_+
\right\}
&=&
2\tilde{\rm T}{\rm r}\left\{
\hat{\Gamma}\hat{\rho}_{\rm W}^2
\right\}\;,
\end{eqnarray}
we finally obtain the entropy production
\begin{eqnarray}
\dot{S}_{\rm lin,W}
&=&
\frac{4}{\hbar}(2\pi\hbar)^N
\left\{
{\rm Tr}^\prime\left[\hat{\Gamma}\hat{\rho}_{\rm S}^2(t)\right]
-{\rm Tr}^\prime\left[\hat{\Gamma}\hat{\rho}_{\rm S}(t)\right]
{\rm Tr}^\prime\left[\hat{\rho}_{\rm S}^2(t)\right]
\right\}\;.
\label{eq:SlinW}
\end{eqnarray}

Within the quantum-classical framework,
we can also introduce a non-Hermitian linear entropy~as
\begin{eqnarray}
S_{{\rm lin},W}^{\rm NH}&=&
1-(2\pi\hbar)^N
\tilde{\rm T}{\rm r}\left[\hat{\rho}_{\rm W}(X,t)\hat{\Omega}_{\rm W}(X,t)\right].
\label{eq:s2WNH}
\end{eqnarray}

The entropy production is given by
\begin{eqnarray}
\dot{S}_{{\rm lin},W}^{\rm NH}
&=&
-2\frac{(2\pi\hbar)^N}{Z_{\rm W}}
\tilde{\rm T}{\rm r}\left[
\hat{\Omega}_{\rm W}
\frac{\partial\hat{\Omega}_{\rm W}}{\partial t}
\right]
-
\frac{2(2\pi\hbar)^N}{\hbar}
\tilde{\rm Tr}\left[\hat{\Gamma}\hat{\Omega}_{\rm W}\right]
\tilde{\rm Tr}\left[\hat{\rho}_{\rm W}^2\right],
\end{eqnarray}
where we have defined
\begin{eqnarray}
Z_{\rm W}&=&\tilde{\rm T}{\rm r}\left[\hat{\Omega}_{\rm W}(X,t)\right].
\end{eqnarray}

In the following, we will use
\begin{eqnarray}
\dot{Z}_{\rm W}&=&
-\frac{2}{\hbar}\tilde{\rm T}{\rm r}\left[\hat{\Gamma}\hat{\Omega}_{\rm W}(X,t)\right].
\end{eqnarray}

In order to calculate $\tilde{\rm T}{\rm r}
[\hat{\Omega}_{\rm W} \partial\hat{\Omega}_{\rm W}/\partial t]$,
we are led to consider the following identities:
\begin{eqnarray}
\tilde{\rm T}{\rm r}\left[
{\cal B}_{ab}\hat{\Omega}_{\rm W}(\nabla_a\hat{H}_W)(\nabla_b\hat{\rho}_W)
\right.
-
\left.
{\cal B}_{ab}\hat{\Omega}_{\rm W}(\nabla_a\hat{\rho}_W)(\nabla_b\hat{H}_W)
\right]
&=&0
\;, \\
\tilde{\rm T}{\rm r}\left\{
\hat{\Omega}_{\rm W}\left[\hat{\Gamma},\hat{\Omega}_{\rm W}\right]_+
\right\}
&=&
2\tilde{\rm Tr}^\prime\left[\hat{\Gamma}\hat{\Omega}_{\rm S}^2\right].
\end{eqnarray}

Finally, we obtain
\begin{eqnarray}
\dot{S}_{\rm lin,W}^{\rm NH}
&=&
\frac{4(2\pi\hbar)^N}{\hbar}
{\rm Tr}^\prime\left[
\hat{\Gamma}\hat{\rho}_{\rm S}\hat{\Omega}_{\rm S}
\right]
-
\frac{2(2\pi\hbar)^N}{\hbar}
{\rm Tr}^\prime\left[\hat{\Gamma}\hat{\Omega}_{\rm S}\right]
{\rm Tr}^\prime\left[\hat{\rho}_{\rm S}^2\right].
\label{eq:SlinWNH}
\end{eqnarray}

\subsection*{Quantum-Classical Linear Entropy Production
and Constant Decay Operator}

When the decay operator $\hat\Gamma$ is given by Equation~(\ref{eq:Gammamodgau}),
Equation~(\ref{eq:qcdotOm}) becomes
\begin{eqnarray}
\frac{\partial}{\partial t}\hat{\Omega}_{\rm W}(X,t)
=&-&\frac{i}{\hbar}
\left[\hat{H}_{\rm W},\hat{\Omega}_{\rm W}(X,t)\right]_-
+
\frac{1}{2} \hat{H}_{\rm W}\overleftarrow{\nabla} \cdot
\mbox{\boldmath${\cal B}$} \cdot
\overrightarrow{\nabla} \hat{\Omega}_{\rm W}(X,t)
\nonumber\\
&-&
\frac{1}{2} \hat{\Omega}_{\rm W}(X,t) \overleftarrow{\nabla} \cdot
\mbox{\boldmath${\cal B}$} \cdot
\overrightarrow{\nabla} \hat{H}_{\rm W}
-
\gamma_0\hat{\Omega}_{\rm W}(X,t) \;, 
\label{eq:qcdotOm-Gamconst}
\end{eqnarray}
and Equation~(\ref{eq:dotqcTrace}) becomes
\begin{equation}
\frac{d}{dt}\tilde{\rm T}{\rm r}\left[\hat{\Omega}_{\rm W}(X,t)\right]
=
-\gamma_0
\tilde{\rm T}{\rm r}\left[\hat{\Omega}_{\rm W}(X,t)\right]
\;.
\label{eq:dotqcTrace-Gamconst}
\end{equation}

Upon choosing the initial condition 
$\tilde{\rm T}{\rm r}\left[\hat{\Omega}_{\rm W}(X,0)\right]=1$,
Equation~(\ref{eq:dotqcTrace-Gamconst}) has the solution
\begin{equation}
\tilde{\rm T}{\rm r}\left[\hat{\Omega}_{\rm W}(X,t)\right]
=
\exp\left[-\gamma_0 t\right]\;,
\end{equation}
which is analogous to Equation~(\ref{eq:trOmega}).
Equation~(\ref{eq:qcdotrhoW}) becomes
\begin{equation}
\begin{array}{ll}
\frac{\partial}{\partial t}\hat{\rho}_{\rm W}(X,t)
=-\frac{i}{\hbar}
\left[\hat{H}_{\rm W},\hat{\rho}_{\rm W}(X,t)\right]_-
+
\frac{1}{2} \hat{H}_{\rm W}\overleftarrow{\nabla} \cdot
\mbox{\boldmath${\cal B}$} \cdot \vspace{4pt}
\overrightarrow{\nabla} \hat{\rho}_{\rm W}(X,t) \\ 
~~~~~~~~~~~~~~~~~~~~~~~-\frac{1}{2} \hat{\rho}_{\rm W}(X,t) \overleftarrow{\nabla} \cdot
\mbox{\boldmath${\cal B}$} \cdot
\overrightarrow{\nabla} \hat{H}_{\rm W}
\;. \\
\label{eq:qcdotrhoW-Gamconst}
\end{array}
\end{equation}

 \vspace{-5pt}

Equation~(\ref{eq:qcdotrhoW-Gamconst}) shows that, in the case
considered, the normalised density matrix $\hat{\rho}_{\rm W}(X,t)$
is not influenced by $\hat\Gamma$, so this
evolves according to the unitary quantum-classical dynamics
{that were first derived in \cite{kapra-cicco}.}

We also have that
Equations~(\ref{eq:SlinW}) and~(\ref{eq:SlinWNH}) become
\begin{eqnarray}
\dot{S}_{\rm lin,W}
&=&
2\gamma_0(2\pi\hbar)^N
\left\{
\tilde{\rm T}{\rm r}^\prime\left[\hat{\rho}_{\rm W}^2(X,t)\right]
-
\tilde{\rm T}{\rm r}\left[\hat{\rho}_{\rm W}^2(X,t)\right]
\right\}=0\;, \label{eq:SlinW-Gamconst} \\
\dot{S}_{\rm lin,W}^{\rm NH}
&=&
\frac{4(2\pi\hbar)^N}{\hbar}
\tilde{\rm T}{\rm r}\left[
\hat{\Gamma}\hat{\rho}_{\rm W}(X,t)\hat{\Omega}_{\rm W}(X,t)
\right]
-
\frac{2(2\pi\hbar)^N}{\hbar}
\tilde{\rm T}{\rm r}\left[\hat{\Gamma}\hat{\Omega}_{\rm W}(X,t)\right]
\tilde{\rm T}{\rm r}\left[\hat{\rho}_{\rm W}^2(X,t)\right]
\nonumber\\
&=&
(2\pi\hbar)^N\gamma_0e^{\gamma_0 t}
\tilde{\rm T}{\rm r}\left[\hat{\Omega}_{\rm W}^2(X,t)\right]\;.
\label{eq:dotSlinWNH-Gamconst}
\end{eqnarray}

In order to evaluate Equation~(\ref{eq:dotSlinWNH-Gamconst}), we need
to calculate $\tilde{\rm T}{\rm r}[\hat{\Omega}_{\rm W}^2(X,t)]$.
From Equation~(\ref{eq:qcdotOm-Gamconst}), we get
\begin{eqnarray}
\frac{\partial}{\partial t}
\tilde{\rm T}{\rm r}\left[\hat{\Omega}_{\rm W}^2(X,t)\right]
&=&
-2\gamma_0
\tilde{\rm T}{\rm r}\left[\hat{\Omega}_{\rm W}^2(X,t)
\right]\;,
\\
\tilde{\rm T}{\rm r}\left[\hat{\Omega}_{\rm W}^2(X,t)\right]
&=&
\tilde{\rm T}{\rm r}\left[\hat{\Omega}_{\rm W}^2(X,0)\right]
\exp\left[-2\gamma_0 t\right]\;.\label{eq:TrOmW-t}
\end{eqnarray}

Upon substituting Equation~(\ref{eq:TrOmW-t}) into Equation~(\ref{eq:dotSlinWNH-Gamconst})
and integrating, we finally obtain
\begin{equation}
S_{\rm lin,W}^{\rm NH}=(2\pi\hbar)^N
\tilde{\rm T}{\rm r}\left[\hat{\Omega}_{\rm W}^2(X,0)\right]
\left(1-e^{-\gamma_0t}\right)\;.
\label{eq:SlinWNH-t}
\end{equation}

Analogously to the pure quantum case, the rate of production
of the quantum-classical entropy in Equation~(\ref{eq:SlinWNH-t})
monitors the flow of information associated with the decay of the purity
of the quantum-classical non-Hermitian system (for positive 
$\gamma_0$).

\section{Conclusions}
\label{sec:concl}

In this paper, we have shown that it is possible to define
meaningful entropy functionals for open quantum systems
described by non-Hermitian Hamiltonians.
In particular, a non-Hermitian generalisation of
the von Neumann entropy, which is able to signal
the loss of information of the quantum subsystem,
requires both the normalised and the non-normalised density matrix:
this entropy can be defined as the normalised average of the logarithm
of the non-normalised density matrix~\cite{sk-S}.

Motivated by the Wigner representation of quantum mechanics,
we have also introduced the non-Hermitian generalisation of
the linear entropy, defined as one minus
the normalised average of the square of the non-normalised
density matrix. Through the analytical solution of the 
 {basic}
case of a constant decay operator, we have shown
that the non-Hermitian linear entropy is able to describe
the loss of purity of the quantum subsystem.
This is true both for pure non-Hermitian subsystems
as well as for non-Hermitian subsystems embedded in 
a classical environment.
 {It is worth repeating that even basic models with
constant decay operators are interesting when one adds
the additional level of complexity provided by the classical-like
environment represented by means of the partially Wigner-transformed
Hermitian part of the Hamiltonian, as in the case of
a light-emitting quantum dot coupled to
an energy-absorbing optical guide in a classical environment.}

The results obtained so far~\cite{ks,kostya-purity,sk-corr,sk-S}
show that the correct description of the dynamics and of the
information flow of systems
described by non-Hermitian Hamiltonians needs the use
of both the normalised and non-normalised density matrix.
In this way, reasonable entropy functionals 
can be introduced.
On conceptual grounds, one might have expected that the
foundation of the non-Hermitian theory on the \emph{normalised}
density matrix alone would hide the interesting effects
arising from the coupling to the probability sinks or sources.~As a matter of fact, the density matrix $\hat\rho$ is constrained
to be normalised in order to be able to define correctly
(normalised) statistical averages. However, such a procedure
inevitably masks the flow of information:
it is as if one would like to study the motion of a body by 
choosing the frame of reference that moves together with the
body itself.
On~the contrary, the flow of information in systems
modelled with non-Hermitian Hamiltonians
can be \emph{solely}
captured through the use of the \emph{non-normalised}
density matrix.

 {We} hope that
the results discussed in this paper
may be a first step toward
{a rigorous analysis of the quantum information flow in}
 systems
with non-Hermitian Hamiltonians, after removing the constraints
of PT-symmetry~\cite{gardas}.


\acknowledgments{Our research has been self-funded through the salaries
received from the University of Messina. We are very grateful to
Professor Ignazio Licata who encouraged and supported us through the
waiver of the publications costs in the journal \emph{Entropy}.}


\end{document}